\documentclass[11pt]{article}
\usepackage{amsfonts}

\title{Boyer-Finley equation and systems of hydrodynamic type}

\def\be{\begin{equation}}
\def\ee{\end{equation}}
\def\beqs{\begin{displaymath}}
\def\eeqs{\end{displaymath}}
\def\beqn{\begin{eqnarray}}
\def\eeqn{\end{eqnarray}}
\def\cp{\mathbb {CP}^1}
\def\g{\gamma}
\def\l{\lambda}

\def\R{\mathbb {R}}

\def\a{\alpha}
\def\p{\partial}
\def\d{\partial}

\def\la{\label}
\def\f{\frac}

\def\L{{\cal L}}

\date{}
\newlength{\dinwidth}
\newlength{\dinmargin}
\def\ben{\begin{displaymath}}
\def\een{\end{displaymath}}
\def\baa{\begin{eqnarray}}
\def\eaa{\end{eqnarray}}

\def\ba{\begin{array}}
\def\ea{\end{array}}

\def\a{\alpha}
\def\g{\gamma}

\def\l{\lambda}

\def\phi{\varphi}

\def\CP1{{\mathbb C}P^1}

\def\la{\label}

\def\f{\frac}
\def\L{{\cal L}}

\def\ctg{{\rm cot}}

\newtheorem{remark}{Remark}
\newtheorem{theorem}{Theorem}

\begin{document}
\maketitle
\begin{center}
\end{center}
\author{\begin{center}
{\Large Ferapontov E.V.},\bigskip \\
Department of Mathematical Sciences \\
    Loughborough University \\
    Loughborough, Leicestershire LE11 3TU \\
    United Kingdom \\
    e-mail: {\tt E.V.Ferapontov@lboro.ac.uk}\bigskip\bigskip\\
{\Large Korotkin D.A., Shramchenko V.A.}\bigskip\\
Department of Mathematics and Statistics\\
Concordia University\\
7141, rue Sherbrooke Ouest\\
Montr\'eal, Qu\'ebec H4B 1R6\\
Canada\\
 e-mails: {\tt korotkin@mathstat.concordia.ca}\\
  {\tt vasilisa@alcor.concordia.ca}\\
\end{center}}
\vskip1.0cm

{\bf Abstract.} We reduce Boyer-Finley equation  to a family of compatible systems of hydrodynamic type,
with characteristic speeds expressed in terms of spaces of rational
functions.
The systems of hydrodynamic type
 are then solved by the generalized hodograph method, providing  
solutions of the Boyer-Finley equation including functional parameters.

\vskip0.7cm
In this paper we construct solutions of the dispersionless non-linear
PDE -- the Boyer-Finley equation (self-dual Einstein equation with a
Killing vector), 
\be
U_{xy}=(e^U)_{tt},
\la{1}
\ee
via reduction to a family of compatible systems of hydrodynamic type.

This equation was actively studied during last twenty years by
many authors; we just mention works \cite{BoyFin82,CalTod01,Ward90,Save92, 
MaShWi,Dun,Winternitz, Kric89,Kric94,ManAlo,GuMaAl}. 
So far the most general scheme of the construction of its solutions was
developed in
\cite{Kric89,Kric94}. In these works solutions of  the Boyer-Finley 
equation were derived by averaging 
an appropriate two-point Baker-Akhiezer function in genus zero which 
corresponds to the
two-dimensional Toda lattice equations, the underlying  Riemann surface deforming 
according to the Whitham equations. 
Some particular solutions of the Boyer-Finley equation were constructed in 
\cite{CalTod01,Ward90,Dun,MaShWi}; their relationship to the solutions of
\cite{Kric89,Kric94} remains unclear. 

The goal of the present  paper is to give an alternative scheme of solving 
the Boyer-Finley equation. Namely, we consider reductions of this
equation to multi-component systems of hydrodynamic type in the spirit of 
\cite{GibTsa96, GibTsa99}, see also \cite{Lei, ManAlo, GuMaAl}; the
equations for charactersitic speeds of these systems are solved in
terms of rational branched coverings. 
The  systems of hydrodynamic type  are then
 solved by the generalized  hodograph method \cite{Tsar90}.

Now we describe a way to find solutions of the Boyer-Finley equation. 
First we ignore the condition of reality of the function $U$ and 
construct complex-valued solutions of equation (\ref{1}); 
then we formulate restrictions on the parameters  which guarantee the reality of $U.$

Let us assume $U$ to be a function of $L$ variables 
$\l_1,\dots,\l_L$, where   $\l_m(x,y,t)$ (``Riemann invariants") satisfy a 
pair of systems of hydrodynamic type
\begin{equation}
\p_x\l_m=V_m(\{\l_k\})\p_t\l_m\;,\hspace{0.7cm}
\p_y\l_m=W_m(\{\l_k\})\p_t\l_m\;.
\label{HD2}
\end{equation}
A direct substitution of the function $U(\l_1,\dots,\l_L)$ into 
the Boyer-Finley equation implies the algebraic relation among the
functions $U\;,\;V_m$ and $W_m$,
\be
V_m W_m = e^U,
\la{UVW2}
\ee
along with the following differential equations:
\be
\p_m\p_n U (V_m W_n + V_n W_m) =2 \p_n\p_m (e^U)\;,\hskip0.7cm m\neq
n,
\la{VWBF1}
\ee
\be
\f{\p_m V_n}{V_m-V_n} = \f{\p_m W_n}{W_m-W_n}\;,\hskip0.7cm m\neq n,
\label{VWBF}
\ee
where $\p_m\equiv\p/\p\l_m$.

Relations (\ref{UVW2}) allow one to parameterize the functions $V_m$ and $W_m$
by a new set of variables $\phi_m(\{\l_k\}) $ as follows:
\be
V_m= \exp\left\{2i\phi_m +\f{U}{2}\right\}\;,\hskip0.7cm 
W_m= \exp\left\{-2i\phi_m +\f{U}{2}\right\}\;.
\la{VmWm}
\ee
In terms of these variables, equations  (\ref{VWBF1}) and  (\ref{VWBF}) 
take the form
\be
\p_m\p_n U=-\f{\p_m U\p_n U}{2\sin^2 (\phi_m-\phi_n)}\;,
\hskip0.7cm
\p_m\phi_n=\f{1}{4}\ctg(\phi_n-\phi_m) \p_m U\;,
\la{mainsys}
\ee
where $m\neq n\;$.

A class of solutions of this system is related to the space of rational functions in the following way. Consider a rational function
\be
R(\mu)=\mu+\sum^{N-1}_{k=1}\frac{a_k}{\mu-b_k}\;,\hspace{0.3cm}\mu\in\cp.
\label{Rofmu}
\ee
In application to Benney's hierarchy, functions of this form  were first introduced in \cite{KupMan77}.
The equation
\be
\l=R(\mu)
\label{covering}
\ee
defines an $N-$sheeted covering $\L$ of the $\l-$sphere. 
A point $P\in\L$ is a pair of complex numbers, $P=(\l,\mu)$. 
We consider the generic case when the function $R(\mu)$ has 
$2N-2$ non-coinciding finite critical points, i.e., the equation
\ben
R^\prime(\mu)=0\;
\een
has $2N-2$ distinct roots $\mu_1,\dots,\mu_{2N-2}.$ The corresponding 
critical values,
\be
\l_n=R(\mu_n),\hspace{0.3cm}n=1,\dots,2N-2,
\ee
are projections onto the $\l-$sphere of the branch points of the covering $\L$ (we denote branch points by $P_n=(\l_n,\mu_n)$; all of them are simple as a corollary of non-coincidence of $\mu_n$ for different $n$). An additional condition we impose on the function $R(\mu)$ is that all $\l_n$ are different.
Now, observe that the number of parameters of the rational function 
(\ref{Rofmu}) is equal to the number of branch points; therefore, we can take $\l_1,\dots,\l_{2N-2}$ as local coordinates on the space of rational functions.
It was shown in \cite{KokKor} that the critical points  $\{\mu_m\}$ of the rational function $R(\mu)$ depend on $\{\l_n\}$ in the following way:
\baa
\frac{\d\mu_m}{\d\l_n}=\frac{\beta_n}{\mu_n-\mu_m}\;;\hspace{0.5cm}\frac{\d\beta_m}{\d\l_n}=\frac{2\beta_n\beta_m}{(\mu_n-\mu_m)^2}\;,\hspace{0.5cm} m\neq n\;.
\label{MuAndBetaDiff}
\eaa
These equations appeared also in \cite{GibTsa96, GibTsa99} in the theory of hydrodynamic reductions
of Benney's moment equations.
The inverse function $\mu(P)=R^{-1}(P)$ is defined on the covering $\L$ . 
As a function of $\{\l_n\}$, it satisfies the system of differential equations 
\cite{KokKor}:
\be
\frac{\d\mu}{\d\l_n}=\frac{\beta_n}{\mu_n-\mu}\;.
\label{mudiff}
\ee
Let us now choose two points $Q_1$ and $Q_2$ on the covering $\L$ such 
that their projections $\l(Q_1)$ and $\l(Q_2)$ onto the $\l-$sphere do not depend on $\{\l_n\}$. Then, consider the following function:
\be
\g(P)=\frac{1}{2\pi i}\log\frac{\mu(P)-\mu(Q_1)}{\mu(P)-\mu(Q_2)}\;,
\label{gamma}
\ee
which maps $\L$ onto a cylinder. We shall be interested in the 
images $\{\g_m\}$ of branch points under this map as functions of $\{\l_n\}.$ 
In the sequel we denote $\mu(Q_1)$ by $\kappa_1$ and $\mu(Q_2)$ by $\kappa_2\;$.  
According to (\ref{mudiff}), they satisfy the equations
\be
\frac{\d\kappa_j}{\d\l_n}=\frac{\beta_n}{\mu_n-\kappa_j}\;,\hskip1.0cm j=1,2\;.
\la{eqkappa}
\ee
From the expression (\ref{gamma}) for $\g(P_m)$ we have
\be
\g_m=\frac{1}{2\pi i}\log\frac{\mu_m-\kappa_1}{\mu_m-\kappa_2}\;.
\label{gamman}
\ee
Differentiation of this relation with respect to $\l_n$ using 
(\ref{MuAndBetaDiff}), (\ref{gamma}) and (\ref{eqkappa}) gives
\baa
\frac{\d\g_m}{\d\l_n}=
\frac{1}{2\pi i}\frac{\beta_n}{\mu_n-\mu_m}\left[ \frac{1}{\mu_m-\kappa_1}-\frac{1}{\mu_m-\kappa_2} \right]\;.
\eaa
If we now express $\mu_n$ and $\mu_m$ in terms of $\gamma_n\;,\;\g_m$ 
from (\ref{gamman}) and set
\be
\alpha_n=-\frac{\beta_n}{4\pi^2}
\left[ \frac{1}{\mu_m-\kappa_1}-\frac{1}{\mu_m-\kappa_2} \right]^2\;,
\label{alphas}\ee
 we arrive at the following system of differential equation for the functions $\g_m(\{\l_n\})\;$:
\be
\frac{\d\g_m}{\d\l_n}=-\pi\alpha_n\Big(\ctg\pi(\g_m-\g_n)+\ctg\pi\g_n\Big)\;,\hspace{0.5cm} m\neq n\;.
\label{gammas}\ee
Similarly, the functions $\alpha_n$ satisfy the following equations:
\be
\frac{\d\alpha_n}{\d\l_m}=2\pi^2\frac{\alpha_n\alpha_m}{\sin^2\pi(\g_n-\g_m)}\;,\hspace{0.5cm} m\neq n\;.
\label{alphasdiff}
\ee

It turns out that a simple transformation allows one to construct solutions of system (\ref{mainsys})
from the set of functions $\g_m$ and $\a_m\;$. Namely, the system 
of equations (\ref{gammas}), (\ref{alphasdiff}) coincides with system 
(\ref{mainsys}) if rewritten in terms of the new variables $U$ and $\phi_n$ such that
\be
\frac{\d U}{\d\l_m}=-4\pi^2\alpha_m\
\label{defU}\ee
and
\be
\phi_n=\pi(\g_n+\psi)\;,
\label{defphi}\ee
where 
\be
\frac{\d \psi}{\d\l_m}=\pi\alpha_m\ctg\pi\g_m\;.
\label{deff}\ee
The existence of functions $U$ and $\psi$ is provided by the 
compatibility conditions,
\be
\f{\d}{\d\l_n} \a_m =
\f{\d}{\d\l_m} \a_n
\ee
and
\be
\f{\d}{\d\l_n} \left(\a_m \ctg\pi\g_m\right)=
\f{\d}{\d\l_m} \left(\a_n \ctg\pi\g_n\right)\;,
\ee
which follow from (\ref{alphasdiff}) and (\ref{gammas}).

Ultimately, formulae (\ref{VmWm})  determine $\{V_m\}$ and $\{W_m\}$ 
 as functions of $\{\l_k\}$. 
In order to obtain a solution of the Boyer-Finley equation (\ref{1}) we 
need  $U$ as an explicit function of $x$, $y$ and $t$, that is, we need to 
solve the system of hydrodynamic type (\ref{HD2}).
The tool which is usually used for this purpose is the generalized 
hodograph method \cite{Tsar90}.
Instead of solving (\ref{HD2}), we find a smooth  solution 
$\Lambda(x, y, t)=\left(\l_1,\dots,\l_L\right)$ of the following system
\be
\Phi_m(\Lambda)=t+V_m(\Lambda)x+W_m(\Lambda)y,
\label{hod}
\ee
where the functions $\{\Phi_m\}$ satisfy the linear system
\be
\frac{\d_m\Phi_n}{\Phi_m-\Phi_n}=\frac{\d_mV_n}{V_m-V_n}=\frac{\d_mW_n}{W_m-W_n}\;,\hspace{0.3cm} m\neq n\;.
\label{PhiVW}
\ee
To see that an implicit solution (\ref{hod}) for $\{\l_m(x,y,t)\}$ 
indeed satisfies (\ref{HD2}), one  needs to differentiate (\ref{hod}) with respect to $x\;,y$ and $t$ \cite{Tsar90}.
 
To be able to use this method we need to construct functions $\Phi_m\;$, i.e. we need to solve for $\Phi_m$ the system
\be
\frac{\d_m\Phi_n}{\Phi_m-\Phi_n}=\frac{\d_mV_n}{V_m-V_n}\;,\hspace{0.3cm} m\neq n\;.
\label{temp}\ee

Observe that for $m\neq n$
\be
\frac{\d_nV_m}{V_m-V_n}=-\frac{\pi^2\alpha_n}{\sin^2\pi(\g_m-\g_n)}\;;
\ee
this is a simple corollary of definitions of $V_m$ and $W_m$ and equations (\ref{defU}), (\ref{mainsys}).

Then the following functions satisfy equations (\ref{temp}):
\be
\Phi_m =\pi^2\oint_l  \f{H(\l) d\g}{\sin^2\pi(\g-\g_m)}\;,
\la{Phim}
\ee
where $l$ is an arbitrary closed contour on the branched covering   $\L$
such that its projection on the $\l$-plane does not depend on the
branch points $\{\l_n\}$ and such that $P_m\notin l$ for all $m$
; $H(\l)$ is an arbitrary function on $l$ 
independent of $\{\l_n\}$.  
The proof of this fact is a simple calculation using equations 
(\ref{gammas}), (\ref{mudiff}) and the link (\ref{gamma}) between $\g$ and $\mu.$

Note that in this framework we can fix positions of branch points 
$\{\l_{L+1},\dots,\l_{2N-2}\}$ and consider the dependence of all functions on the remaining set of variables $\{\l_1,\dots,\l_L\},$ where $L\leq2N-2.$

The following theorem summarizes our construction of solutions of
the Boyer-Finley equation.
\begin{theorem}\la{theosol}
Let functions $\a_m(\{\l_n\})$ and $\g_m(\{\l_n\}),$ $\;m,n=1,\dots,L$, \
 $L\leq 2N-2$, be associated with an $N$-sheeted 
branched covering as described above. Define the 
potentials $U(\{\l_n\})$ and $\psi(\{\l_n\})$ to be solutions of the
following system of equations:
\be\la{Ulm1}
\f{\p U}{\p\l_m}=-4\pi^2 \a_m\;, \hspace{0.5cm} m=1,\dots,L\;;
\ee
\be\la{flm1}
\f{\p \psi}{\p\l_m}=\pi \a_m\ctg \pi\g_m\;,\hspace{0.5cm} m=1,\dots,L\;.
\ee
Let the $(x,y,t)-$dependence of branch points $\l_n,\;n=1,\dots,L$, be governed by the
following system of $L$ equations,
\be
\pi^2 \oint_l \f{H(\l)d\g}{\sin^2\pi (\g-\g_m)} = t+ x\, V_m+ y\,
W_m\;,\hskip0.7cm m=1,\dots,L\;, 
\la{imlic}
\ee
where 
\be
V_m = e^{2\pi i (\g_m+\psi) +U/2}\;,\hskip0.7cm 
W_m = e^{-2\pi i (\g_m+\psi) +U/2}\;;
\la{VWmm}
\ee
$l$ is an arbitrary $\{\l_n\}$-independent contour on $\L$ such that all $P_m\notin l$;
$H(\l)$ is an arbitrary summable $\{\l_n\}$-independent function on $l$.

Then the function $U(\{\l_n(x,y,t)\})$ satisfies the Boyer-Finley equation (\ref{1}).
\end{theorem}

\begin{remark}\rm
If an $N$-sheeted rational branched covering $\L$ with two marked points $Q_1\;,\;Q_2$ is fixed, the solution of
the Boyer-Finley equation constructed according to this theorem is defined
by 

(a) a functional parameter $H(\l)$ and 

(b) a number $L\leq 2N-2$, which has a 
meaning of the number of components 
$\l_m$ satisfying systems of hydrodynamic type (\ref{HD2}) with 
characteristic speeds  (\ref{VWmm}). 

The application of theorem \ref{theosol} in practice requires calculation
of quadratures (\ref{Ulm1}) and (\ref{flm1}); besides that, one needs
to resolve
implicit relations (\ref{imlic}) to find the dependence of $\l_m$ on $(x,y,t)$.
\end{remark}

So far we were dealing with complex solutions of the Boyer-Finley equation
(\ref{1}); it is easy to formulate conditions on the
parameters of our solutions which provide the reality of the function $U$.

Let us assume the function $R(\g)$ to satisfy the ``reality condition"
\be
\overline{R(\bar{\g})}=R(\g).
\label{Rreality}
\ee
Then
the branch covering $\L$ is invariant with respect to the antiholomorphic involution $\tau\;$, which acts on the points $(\l,\mu)$ of the covering $\L$
as follows:
\be
\tau:\;(\l,\mu)\rightarrow(\bar{\l},\bar{\mu}).
\label{involution}\ee
Assume also that both points $Q_1,\;Q_2$ are invariant with respect to $\tau$, 
i.e., $\kappa_{1,2}\in\R$. 
Let also the contour $l$ be invariant with respect to the involution 
and the function $H(P)$ satisfy the relation $H(P)=-\overline{H(P^\tau)}$.
Then one can choose the constants of integration in (\ref{Ulm1}) and
 (\ref{flm1}) such that the
solution $U(x,y,t)$ of the Boyer-Finley equation given by theorem \ref{theosol} is real.

Indeed, the invariance of the covering $\L$ with respect to $\tau$ means 
that all $\l_m$ are either real or form conjugate pairs; 
the same holds for the set $\{\mu_m\}$. The expression (\ref{gamma}) for the map $\g$ implies
\be
\overline{\g(P^\tau)}=-\g(P)\;,
\la{realgamma}
\ee
therefore, all $\g_m\equiv \g(P_m)$ are either imaginary,
 $\overline{\g_m}=-\g_m$,
or form anti-conjugate pairs.
Applying complex conjugation to both sides of equation (\ref{gammas}), we find that $\a_m$ are either real ($\a_m\in\R$ if $\l_m\in\R$) or form
conjugate pairs, $\a_m=\overline{\a_n}$, if $\l_m=\overline{\l_n}\;$. This
readily implies that one can choose the integration constant in the
definition (\ref{Ulm1}) of  potential $U$ in such a way that $U$ is a real
function of $\{\l_m\}$.

For completeness, we should also check that the reality condition does
not contradict the solvability of system (\ref{imlic}).
Assume, for simplicity, that all $\l_m$ are real, i.e., all $\g_m$ are
imaginary and all $\a_m$ are real. Then the potential function 
$\psi$ solving system (\ref{flm1}) 
can be chosen to be imaginary, and both $V_m$ and $W_m$ (\ref{VWmm})
are real. Together with $\overline{H(P^\tau)}=- H(P)$ and
(\ref{realgamma}), it implies that both sides of equations
(\ref{imlic}) are real. Therefore, system (\ref{imlic}) gives $L$ real
equations for $L$ real variables $\{\l_m\}$, and generically has solutions.

Similar consideration applies when some $\l_m$'s form conjugate
pairs; in this case the corresponding equations (\ref{imlic}) will be
conjugate to each other, and the number of real equations will again
coincide with the number of real variables.

{\bf Acknowledgements} We thank Tamara Grava and Yavuz Nutku for
useful discussions.

\end{document}